\documentclass[12pt]{iopart}
\usepackage{iopams}
\begin{document}
\title{ Energy momentum flows for the massive vector field}
\author{George Horton and Chris Dewdney}
\address{Division of Physics, University of Portsmouth. Portsmouth PO1 2DT. England}

\begin{abstract}
We present a causal trajectory interpretation for the massive vector
field, based on the flows of rest energy and a conserved density
defined using the time-like eigenvectors and eigenvalues of the
stress-energy-momentum tensor. This work extends our previous work
which used a similar procedure for the scalar field. The massive,
spin-one, complex vector field is discussed in detail and solutions
are classified using the Pauli-Lubanski spin vector. The flows of
energy-momentum are illustrated in a simple example of standing
waves in a plane.

\end{abstract}
 \maketitle

\section{Introduction}
In previous papers we have given a general method of constructing a
causal trajectory interpretation, in the relativistic domain, for
quantum mechanics \cite{HD1}, and in detail for the Klein-Gordon
equation. This method has then been extended to the many-particle
case (fixed number), which entailed the introduction of a Lorentz
invariant rule for the coordination of the space time points on the
individual particle trajectories which does not entail contradiction
between Lorentz invariance and non-locality \cite{HD3},\cite{HD4}.

In the classical context of general relativity Edelen \cite{Edelen}
has given a natural definition of the flow of rest-energy and a
conserved density in terms of the time-like eigenvectors and
eigenvalues of the stress-energy-momentum tensor. In this paper we
extend the detailed discussion to the massive, spin 1, complex
vector field and give an example of standing waves on a space-like
plane.

In sections two and three we give a general account of the massive
vector field formalism. The classification of the solutions by the
Pauli-Lubanski spin vector is then detailed in sections four, five
and six, which enables a covariant characterisation of intrinsic
spin. The symmetric stress-energy-momentum tensor and its
eigenvectors and eigenvalues are given in sections seven and eight
which requires an extension from the massless maxwellian case. The
two parts of the complex field are shown to give rise to a tetrad of
vectors which each define a time-like two-plane and space-like
two-plane.
\section{The massive vector field}
The massive vector, complex field $\phi^\mu(x)$ has a Lagrangian
density

\begin{eqnarray}\nonumber
\mathcal{L}=&\frac{1}{2}\bar{\mathcal{G}}_{\mu\nu}\mathcal{G}^{\mu\nu}-\frac{1}{2}\bar{\mathcal{G}}_{\mu\nu}(\partial^\mu\phi^\nu-\partial^\nu\phi^\mu)\nonumber\\
& -\frac{1}{2}(\partial^\mu\bar{\phi}^\nu-\partial^\nu\bar{\phi}^\mu)\mathcal{G}_{\mu\nu}\nonumber\\
&+m^2\bar{\phi}_\mu\phi^\mu
\end{eqnarray}

with the derived field equations
\begin{eqnarray}
\mathcal{G}^{\mu\nu}=\partial^\mu\phi^\nu-\partial^\nu\phi^\mu\\
-\partial_\nu \mathcal{G}^{\mu\nu}+m^2\phi^\mu=0
\end{eqnarray}

These are the Proca equations for the vector fields, equivalent to
\begin{eqnarray}
(\Box+m^2)\phi_\mu=0\\
\partial_\mu\phi^\mu=0\label{div}
\end{eqnarray}

The condition in equation (\ref{div}) is, of course, essential to
obtain three independent components of $\phi^\mu$ in a covariant
way, so as to describe a spin one field.

It is sometimes convenient to introduce an auxiliary condition to
ensure the splitting of an arbitrary solution $\phi^\mu$ into a
transverse (spin 1) part and a scalar part (as for Stueckelberg's
Lagrangian). One adds a term
$-\lambda(\partial\cdot\phi)(\partial\cdot\bar{\phi})$ to
$\mathcal{L}$. The field $\phi^T_\mu$ is then divergenceless, where
\begin{equation}
\phi^T_\mu=\phi_\mu+\frac{\lambda}{m^2}\partial_\mu(\partial\cdot\phi)
\end{equation}
The equations of motion will be
\begin{equation}
(\Box+m^2)\phi_\mu-(1-\lambda)\partial_\mu(\partial\cdot\phi)=0
\end{equation}
from which, taking the divergence of both sides one gets
\begin{equation}
[\lambda\Box+m^2](\partial\cdot\phi)=0
\end{equation}
One can then explore, in the usual way, the limits $m\rightarrow
0$, and $\lambda\rightarrow 0$.

For the real vector field and field quantization procedures see
\cite{Itzykson}

The field $\phi_\mu(x)$ will, as a result of the condition given
in equation (\ref{div}) have Fourier components of the form
\begin{equation}
a^\mu(k) e^{-ikx}
\end{equation}
with
\begin{equation}
k_\mu a^\mu=0
\end{equation}
that is $a^\mu$ is space-like and $k_\mu$ is time-like. In a
general wave packet, however, $\phi_\mu$ need not be space-like.

In general one can write \cite{Itzykson}
\begin{equation}
\phi_\mu(x)=\int\frac{d^3k}{2k_0(2\pi)^3}\Sigma_{\lambda=1}^3[a^{(\lambda)}(k)\epsilon^{(\lambda)}_{\mu}(k)e^{-ikx}+
a^{(-\lambda)}(k)\epsilon^{(-\lambda)}_{\mu}(k)e^{+ikx}]
\end{equation}
where $k_0=\sqrt{(k^2+m^2)}$.

The three space-like, orthonormalised vectors
$\epsilon^{(\lambda)}_\mu(k)$ are orthogonal to the time-like
vector $k_\mu$ and
\begin{eqnarray}
\epsilon^{(\lambda)}(k)\cdot\epsilon^{(\lambda')}(k)=\delta_{\lambda\lambda'}\\
\Sigma_\lambda\epsilon^{(\lambda)}_\mu(k)\epsilon^{(\lambda)}_\nu(k)=-\left[g_{\mu\nu}-\frac{k_\mu
k_\nu}{m^2}\right]
\end{eqnarray}

\section{Electromagnetic Interaction}
In order to include the effects of interaction of a charged
massive vector field with an electromagnetic field $F_{\mu\nu}$
with vector potential $A_\mu$ one introduces the gauge covariant
operator
\begin{equation}
D_\mu=\partial_\mu+ieA_\mu
\end{equation}
with
\begin{equation}
[D_\mu,D_\nu]=ieF_{\mu\nu}
\end{equation}
The equations of motion will be
\begin{equation}
-D_\nu \mathcal{G}^{\mu\nu}+m^2\phi^\mu=0
\end{equation}
with
\begin{equation}
\mathcal{G}^{\mu\nu}=D^\mu\phi^\nu-D^\nu\phi^\mu
\end{equation}
The second order wave equation and the new divergence condition
imposed on $\phi^\mu$ require more detailed derivation.
\begin{equation}
D_\mu(D_\nu \mathcal{G}^{\mu\nu})=m^2D_\mu\phi^\mu
\end{equation}
but
\begin{equation}
D_\mu D_\nu=D_\nu D_\mu+ieF_{\mu\nu}
\end{equation}
hence
\begin{eqnarray}\nonumber
D_\mu(D_\nu \mathcal{G}^{\mu\nu})&=&D_\nu(D_\mu \mathcal{G}^{\mu\nu})+ieF_{\mu\nu}\mathcal{G}^{\mu\nu}\\
\nonumber&=&D_\nu(-m^2\phi^\nu)+ieF_{\mu\nu}\mathcal{G}^{\mu\nu}\\
\nonumber&=&m^2(D_\nu\phi^\nu)
\end{eqnarray}
Therefore
\begin{equation}
D_\nu\phi^\nu=\frac{ie}{2m^2}F_{\mu\nu}\mathcal{G}^{\mu\nu}\label{newdiv}
\end{equation}
This last condition, equation (\ref{newdiv}), is the new
divergence condition.

The second order equation is given by
\begin{eqnarray}\nonumber
-D_\nu \mathcal{G}^{\mu\nu}&=&D_\nu(D^\nu\phi^\mu)-D_\nu(D^\mu\phi^\nu)\\
\nonumber&=&(D_\nu
D^\nu)\phi^\mu-D^\mu(D_\nu\phi^\nu)+ieF^\mu_\nu\phi^\nu
\end{eqnarray}
Therefore
\begin{equation}
(D_\nu
D^\nu)\phi^\mu+m^2\phi^\mu+ieF^\mu_\nu\phi^\nu-D^\mu(D_\nu\phi^\nu)=0
\end{equation}
and so
\begin{equation}
(D_\nu
D^\nu)\phi^\mu+m^2\phi^\mu+ieF^\mu_\nu\phi^\nu-\frac{ie}{2m^2}D^\mu(F_{\alpha\beta}\mathcal{G}^{\alpha\beta})=0
\end{equation}

It is of interest to compare the term $ieF^\mu_\nu\phi^\nu$ with the
term that occurs in the second order Dirac equation \cite{Itzykson}
namely,
$-g\frac{e}{2}[i\underline{\alpha}\cdot\underline{E}+\underline{\sigma}\cdot\underline{B}]$,
in the usual notation.

$F^\mu_\nu$ can be expressed in terms of the infinitesimal
generators for spatial rotations and Lorentz boosts
\cite{Schweber}
\begin{eqnarray}
\underline{M}&=&[M_{32},M_{13},M_{21}]\\
\underline{N}&=&[M_{01},M_{02},M_{03}]
\end{eqnarray}
$M_{\mu\nu}$ being an anti symmetric tensor corresponding to
infinitesimal rotations in the $\mu\nu$ plane (from $\mu$ to
$\nu$)

Explicitly

\[
M_1=M_{32}= \left[ \begin{array}{cccc}
0 & 0&0&0 \\
0 & 0&0&0\\
0 &0&0&-1\\
0&0&+1&0\\
\end{array}\right]
\]

\[
M_2=M_{13}= \left[ \begin{array}{cccc}
0 & 0&0&0 \\
0 & 0&0&1\\
0 &0&0&0\\
0&-1&0&0\\
\end{array}\right]
\]

\[
M_3=M_{21}= \left[ \begin{array}{cccc}
0 & 0&0&0 \\
0 & 0&-1&0\\
0 &1&0&0\\
0&0&0&0\\
\end{array}\right]
\]

\[
N_1=M_{01}= \left[ \begin{array}{cccc}
0 & 1&0&0 \\
1 & 0&0&0\\
0 &0&0&0\\
0&0&0&0\\
\end{array}\right]
\]

\[
N_2=M_{02}= \left[ \begin{array}{cccc}
0 & 0&1&0 \\
0 & 0&0&0\\
1 &0&0&0\\
0&0&0&0\\
\end{array}\right]
\]

\[
N_3=M_{03}= \left[ \begin{array}{cccc}
0 & 0&0&1 \\
0 & 0&0&0\\
0 &0&0&0\\
1&0&0&0\\
\end{array}\right]
\]
Since
\[
[F^\mu_\nu]= \left[ \begin{array}{cccc}
0 & E^1&E^2&E^3 \\
E^1 & 0&B^3&-B^2\\
E^2 &-B^3&0&B^1\\
E^3&B^2&-B^1&0\\
\end{array}\right]
\]
one can write
\[
F^\mu_\nu=N_kE^k-M_kB^k
\]
That is
\[
F^\mu_\nu\phi^\nu=e[(iN_k)\cdot E^k-(iM_k)B^k]\phi^\nu
\]
This term compares directly with that in the Dirac equation as given
above.
\section{Classification of Solutions according to the
Pauli-Lubanski Spin Vector $W^\mu$}\label{classification} A
covariant characterization of intrinsic spin can be given in terms
of the Pauli-Lubanski operator $W_\mu$
\[
W_\mu=-\frac{1}{2}i\epsilon_{\mu\nu\rho\sigma}M^{\nu\rho} P^\sigma
\] \cite{Itzykson}. Using the previously defined
infinitesimal rotation operators $M_k,N_k$ one has \footnote{The
factor $i$ is sometimes absorbed into $M_k, N_k$ to give a hermitian
and anti-hermitian operator.}
\begin{eqnarray}\nonumber
W^1=-iP_0M^1-i(P_2N^3-P_3N^2)\\\nonumber
W^2=-iP_0M^2-i(P_3N^1-P_1N^3)\\\nonumber
W^3=-iP_0M^3-i(P_1N^2-P_2N^1)\\\nonumber
W^0=-iP_kM^k\\
\end{eqnarray}
where $P_\mu$ is the usual covariant momentum operator.

The actual hermitian spin operators $\underline{S}$ are not
covariant and take the form \cite{Gursey}
\[
\underline{S}=\frac{1}{m}[\underline{W}-\frac{W_0\underline{P}}{m+P_0}]
\]
and obey the commutation rule
\[
[S_p,S_q]=i\epsilon_{pqr}S_r
\]
and $[S_r,P_\lambda]=0$ as expected for spin operators.

We consider the eigenfunctions of $W^3$ and $W^0$. In the rest
frame for a single plane wave there will, of course, be no
difference between $\underline{S}$ and $\underline{W}$.
\section{Eigenfunctions of $W^3$}\label{ef}
Consider a single plane wave
\[
\epsilon_\mu(k)e^{-ik\cdot x}
\]
with
\[
W^3\left(%
\begin{array}{c}
  \epsilon_0 \\
  \epsilon_1 \\
  \epsilon_2 \\
  \epsilon_3 \\

\end{array}%
\right)e^{-ik\cdot x}
=\lambda\left(%
\begin{array}{c}
  \epsilon_0 \\
  \epsilon_1 \\
  \epsilon_2 \\
  \epsilon_3 \\

  \end{array}%
   \right)e^{-ik\cdot x}
\]
Explicitly
\begin{eqnarray}\nonumber
-ik_2\epsilon_1+ik_1\epsilon_2&=&\lambda\epsilon_0\\
\nonumber
-ik_2\epsilon_0+ik_0\epsilon_2&=&\lambda\epsilon_1\\\nonumber
+ik_1\epsilon_0-ik_0\epsilon_1&=&\lambda\epsilon_2\\\nonumber
0&=&\lambda\epsilon_3
\end{eqnarray}
for which $ \lambda=0, \lambda=\pm\sqrt{k_0^2-k_1^2-k_2^2}$

\textbf{Case {$\lambda=0$}}

$\epsilon_1=k_1, \epsilon_2=k_2, \epsilon_0=k_0$

$\epsilon_3$ is chosen to give

\[
\partial^\mu\epsilon_\mu e^{-ik\cdot x}=0
\]
leading to
\[
\epsilon_3=\frac{m^2+k_3^2}{k_3}
\]

\[
\left(%
\begin{array}{c}
  \epsilon_0 \\
  \epsilon_1 \\
  \epsilon_2\\
  \epsilon_3 \\

\end{array}%
\right)e^{-ik\cdot x}
=\left(%
\begin{array}{c}
  k_0 \\
  k_1 \\
  k_2\\
  \frac{m^2+k^2_3}{k_3} \\

  \end{array}%
   \right)e^{-ik\cdot x}
\]

\textbf{Case {$\lambda=\pm\sqrt{k_0^2-k_1^2-k_2^2}$}}

\[
\frac{\epsilon_1}{\epsilon_0}=\frac{\lambda k_0-ik_1k_2}{\lambda
k_1-ik_0k_2}
\]
\[
\frac{\epsilon_2}{\epsilon_0}=\frac{\lambda k_0+ik_1k_2}{\lambda
k_2+ik_0k_1}
\]

For the special circumstance in which only $k_0$ and $k_3$ are
non-zero one has the two cases

\textbf{Case $\lambda=0$}
\[
\left(%
\begin{array}{c}
  \phi_0 \\
  \phi_1 \\
  \phi_2\\
  \phi_3 \\

\end{array}%
\right)
=\left(%
\begin{array}{c}
  k_0 \\
 0 \\
0\\
  \frac{k_0^2}{k_3} \\

  \end{array}%
   \right)e^{-i(k_0x^0+k_3x^3)}
\]
which is the case of longitudinal polarization.

\textbf{Case {$\lambda=\pm k_0$}}
\[
\left(%
\begin{array}{c}
  \phi_0 \\
  \phi_1 \\
  \phi_2\\
  \phi_3 \\

\end{array}%
\right)
=\left(%
\begin{array}{c}
  0 \\
 1 \\
\pm i\\
 0 \\

  \end{array}%
   \right)e^{-i(k_0x^0+k_3x^3)}
\]
which corresponds with the case of right and left circular
polarization.
\subsection{Standing waves with spin perpendicular to the
direction of propagation}\label{spind} For the case in which only
$k_0$ and $k_1$ are non-zero one has
\begin{eqnarray}\nonumber
\frac{\epsilon_1}{\epsilon_0}=\frac{k_0}{k_1}\\\nonumber
\frac{\epsilon_2}{\epsilon_0}=-i\frac{\lambda}{k_1}
\end{eqnarray}
and

$\epsilon_0=k_1, \epsilon_1=k_0, \epsilon_2=\mp |\lambda|=\mp im$\\
For the case in which only $k_0$ and $k_2$ are non-zero one has

$\epsilon_0=k_2, \epsilon_2=k_0, \epsilon_1=\pm im$

In both cases
\[\partial^\mu\epsilon_\mu e^{-ik\cdot x}=0\]

One can then construct the single frequency state for counter
propagating waves in the $x^1$ and $x^2$ directions with spin in the
third $x^3$ direction. In this case there is no difference between
$S^3$ and $\frac{W^3}{m}$. In the positive spin case one has

\begin{equation}\label{phivector}
\left(%
\begin{array}{c}
  \phi_0 \\
  \phi_1 \\
  \phi_2\\
  \phi_3 \\

\end{array}%
\right)
=\left(%
\begin{array}{c}
  -i(k_1\sin(k_1x_1)+k_2\sin(k_2x_2)) \\
 im\cos(k_2x_2)+k_0 \cos(k_1x_1) \\
-im\cos(k_1x_1)+k_0\cos(k_2x_2)\\
 0 \\

  \end{array}%
   \right)e^{-i(k_0x^0)}
\end{equation}
with $|k_1|=|k_2|$.

\section{Eigenfunctions of $W^0$}
As for $W^3$
\[
W^0\left(%
\begin{array}{c}
  \epsilon_0 \\
  \epsilon_1 \\
  \epsilon_2 \\
  \epsilon_3 \\

\end{array}%
\right)e^{-ik\cdot x}
=\lambda\left(%
\begin{array}{c}
  \epsilon_0 \\
  \epsilon_1 \\
  \epsilon_2 \\
  \epsilon_3 \\

  \end{array}%
   \right)e^{-ik\cdot x}
\]
Explicitly
\begin{eqnarray}\nonumber
-ik_3\epsilon_2+ik_2\epsilon_3&=&\lambda\epsilon_1\\\nonumber
+ik_3\epsilon_1-ik_1\epsilon_3&=&\lambda\epsilon_2\\\nonumber
-ik_2\epsilon_1+ik_1\epsilon_2&=&\lambda\epsilon_3
\end{eqnarray}
which gives
\begin{eqnarray}\nonumber
\lambda=0\\\nonumber
\lambda=\pm\sqrt{k_0^2-m^2}=\pm\sqrt{k_1^2+k_2^2+k_3^2}
\end{eqnarray}
since \begin{equation}\label{e1}
i(\underline{k}\times\underline{\epsilon})=\lambda\underline{\epsilon}
\end{equation}
then
\begin{equation}\label{e2}
(\underline{k}\times\underline{\epsilon})(\underline{k}\times\bar{\underline{\epsilon}})=\lambda^2\underline{\epsilon}\cdot\underline{\bar{\epsilon}}\label{2}
\end{equation}
and
\begin{equation}\label{e3}
\lambda^2\underline{\epsilon}\cdot\underline{\bar{\epsilon}}=(\underline{k}\cdot\underline{k})(\underline{\epsilon}\cdot\underline{\bar{\epsilon}})
-(\underline{k}\cdot\underline{\epsilon})(\underline{k}\cdot\underline{\bar{\epsilon}})
\end{equation}
But, from equation (\ref{e1})
\begin{equation}
\lambda(\underline{k}\cdot\underline{\epsilon})=i\underline{k}\cdot(\underline{k}\times\underline{\epsilon})=0
\end{equation}
Hence $\underline{\epsilon}$ is transverse with respect to
$\underline{k}$, with $\lambda^2=\underline{k}\cdot\underline{k}$
from equation (\ref{e3}) unless
($\underline{k}\times\underline{\epsilon}=0$). Some simple algebra
gives, for $\lambda=0$,
\[
\left(%
\begin{array}{c}
  \epsilon_0 \\
  \epsilon_1\\
  \epsilon_2 \\
  \epsilon_3\\

\end{array}%
\right)
=\left(%
\begin{array}{c}
  \frac{k_0^2-m^2}{k_0} \\
  k_1 \\
 k_2\\
  k_3 \\

  \end{array}%
   \right)
\]
and for $\lambda=\pm\sqrt{\underline{k}\cdot\underline{k}}$, from
equation (\ref{e1}) one easily gets
\[
\frac{\epsilon_1}{\epsilon_3}=\frac{-k_1k_3+i\lambda
k_2}{\lambda^2-k_3^2}
\]
\[
\frac{\epsilon_2}{\epsilon_3}=\frac{-k_2k_3-i\lambda
k_1}{\lambda^2-k_3^2}
\]
\[
\epsilon_0=0
\]
\section{Symmetric stress energy-momentum tensor}
A symmetric tensor can be derived by considering the variation of
the Lagrangian with respect to the metric tensor $g_{\mu\nu}$.
\[
T_{\mu\nu}=\frac{2}{\sqrt{-g}}\left[\frac{\partial\sqrt{-g}\mathcal{L}}{\partial
g^{\mu\nu}}-\partial^\alpha\left(\frac{\partial\sqrt{-g}\mathcal{L}}{\partial\left(\partial^\alpha
g^{\mu\nu}\right)}\right)\right]
\]
In the case being considered there are no terms involving
$\partial^\alpha g^{\mu\nu}$ so that one finds
\[
T_{\mu\nu}=2\frac{\partial\mathcal{L}}{\partial
g^{\mu\nu}}-\mathcal{L}g_{\mu\nu}
\]
Inserting the metric tensor into the previously given
$\mathcal{L}$ one gets
\begin{eqnarray}\nonumber
T_{\mu\nu}=&\bar{\mathcal{G}}_{\mu\alpha}\mathcal{G}^\alpha_\nu+\mathcal{G}_{\mu\alpha}\bar{\mathcal{G}}^\alpha_\nu\\
\nonumber&+m^2(\bar{\phi}_\mu\phi_\nu+\phi_\mu\bar{\phi}_\nu)\\
&+g_{\mu\nu}\left(\frac{1}{2}\left(\bar{\mathcal{G}}_{\alpha\beta}\mathcal{G}^{\alpha\beta}\right)-m^2\bar{\phi}^\alpha\phi_\alpha\right)
\end{eqnarray}
Expressing the complex field in terms of two real fields
$\phi_\mu(1)$ and $\phi_\mu(2)$
\[
\phi_\mu=\phi_\mu(1)+i\phi_\mu(2)
\]
$T_{\mu\nu}$ becomes the sum of the stress-energy momentum tensors
$T_{\mu\nu}(1)$ and $T_{\mu\nu}(2)$ of the above two real fields.
\begin{eqnarray}\nonumber
\frac{1}{2}T_{\mu\nu}(1)=&G_{\mu\alpha}(1)G_\nu^\alpha(1)+m^2\phi_\mu(1)\phi_\nu(1)\\
&+g_{\mu\nu}\left[\frac{1}{4}\left(G^{\alpha\beta}(1)G_{\alpha\beta}(1)\right)-\frac{m^2}{2}\phi^\alpha(1)\phi_\alpha(1)\right]
\end{eqnarray}
and similarly for $\frac{1}{2}T_{\mu\nu}(2)$. Where $G$ is a purely
real antisymmetric field tensor, $\mathcal{G}$ is reserved for the
complex field tensor.

The parts of $T_{\mu\nu}(1)$ and $T_{\mu\nu}(2)$ not depending
explicitly on the mass have the same form as for the massless real
electromagnetic field.

It is known for the massless real (spin 1) field that $T_{\mu\nu}$
can be expressed in terms of a null tetrad of vectors which,
however, are not unique. A natural choice is given by the null
eigenvectors of $G_{\mu\nu}$ (for each of the two real fields)
\cite{Synge1965},\cite{Ruse1936}. Another way of specifying the null
tetrad for the massless real electromagnetic field has been given by
Misner et al \cite{Misner1957} and developed in a recent paper by
Garat \cite{Garat2004}. We give our version of this latter method,
useful in our case, in the appendix to this paper.

The eigenvector equation for either of the two $\it{real}$ fields is
of the form
\[
G_{\mu\nu} X^\nu=\lambda X_\mu
\]
(The ``inscript'' 1,2 is omitted). It then follows from the
antisymmetry of $G_{\mu\nu}$ that the solutions fall into two
groups.

\begin{enumerate}
\item Null eigenvectors defining a time-like two-plane with equal
real eigenvalues of opposite sign.

\item Null eigenvectors defining a space-like two plane with
equal pure imaginary eigenvalues of opposite sign
\end{enumerate}
We denote the null tetrad by $l^\mu, n^\nu, m^\mu, \bar{m}^\mu$
using the Penrose notation \cite{Penrose}
\begin{eqnarray}\nonumber
l^\mu
l_\mu=n^\mu n_\mu=m^\mu m_\mu=\bar{m}_\mu\bar{m}^\mu=0\\
\nonumber l_\mu n^\mu=1\\
m_\mu\bar{m}^\mu=-1
\end{eqnarray}
and all other scalar products vanish.

The metric is given by
\[
g_\mu^\nu=n_\mu l^\nu+l_\mu n^\nu-\bar{m}_\mu
m^\nu-m_\mu\bar{m}^\nu
\]
with the signature $(+,-,-,-)$. Then
\[
G_{\mu\nu}=\lambda(1)[l_\mu n_\nu-n_\mu
l_\nu]+i\lambda(2)[m_\mu\bar{m}_\nu-\bar{m}_\mu m_\nu]
\]
It is easily shown that
\[
\frac{1}{2}G_{\alpha\beta}G^{\alpha\beta}=-(\lambda^2(1)-\lambda^2(2))
\]
\[
G_{\mu\alpha}G^\alpha_\nu=\lambda^2(1)[l_\mu n_\nu+l_\nu
n_\mu]+\lambda^2(2)[m_\mu\bar{m}_\nu+m_\nu\bar{m}_\mu]
\]
or, using the metric tensor
\[
G_{\mu\alpha}G^\alpha_\nu=\left(\lambda^2(1)+\lambda^2(2)\right)\left[l_\mu
n_\nu+l_\nu n_\mu\right]-\lambda^2(2)g_{\mu\nu}
\]
with
\[
\left(\lambda^2(1)+\lambda^2(2)\right)=\frac{1}{4}\left[\left(G_{\alpha\beta}G^{\alpha\beta}\right)^2+\left(G_{\alpha\beta}^*G^{\alpha\beta}\right)^2\right]^\frac{1}{2}
\]
Each real field gives expressions of the same form. However, in
general, the two time-like two-planes defined by $l_\mu,n_\nu$
will be different.

A useful method in finding the eigenfunctions of the total
$T_{\mu\nu}$ is to construct a common time-like two-plane. A unique
prescription can be given following a discussion in \cite{Penrose}
of the properties of the Lorentz transformations. Given any four,
distinct, real null vectors one has a unique time-like two-plane
$\Omega$ which contains

\begin{enumerate}
    \item One vector from each of the planes defined by
    $(l_\mu(1),n_\mu(1))$ and $(l_\mu(2),n_\mu(2))$.
    \item One normal to each of the same two-planes.
\end{enumerate}
If $\Omega$ contains the orthogonal vectors
$\underline{t},\underline{z}$, then the restricted Lorentz
transformation relating the two-planes and conserving $\Omega$ is
a rotation about $\underline{z}$ and a boost along the
$\underline{z}$ axis. Once the tetrads for the two real fields
have been found it is straightforward to determine the overall
eigenvectors.
\section{Classification of the eigenvectors of the real and
imaginary parts of the stress energy-momentum
tensor}\label{secrealT}
\begin{eqnarray}\nonumber\label{Tmunu}
\frac{T_{\mu\nu}}{2}=&G_{\mu\alpha}G^\alpha_\nu+m^2\phi_\mu\phi_\nu\\
\nonumber&+g_{\mu\nu}\left[\frac{1}{4}\left(G^{\alpha\beta}G_{\alpha\beta}\right)-\frac{m^2}{2}\phi^\alpha\phi_\alpha\right]\\
\end{eqnarray}
It is convenient to switch to a tetrad in terms of space and time
components
\begin{eqnarray}
\nonumber l_\mu&=&\frac{1}{\sqrt{2}}(T_\mu+Z_\mu)\\
\nonumber n_\mu&=&\frac{1}{\sqrt{2}}(T_\mu-Z_\mu)\\
\nonumber m_\mu&=&\frac{1}{\sqrt{2}}(X_\mu-iY_\mu)\\
\nonumber \bar{m}_\mu&=&\frac{1}{\sqrt{2}}(X_\mu+iY_\mu)\\
\end{eqnarray}
As we are concerned here with the time-like or space-like
character of the eigenvectors we may just consider a tensor of the
form
\[
k[(T_\mu T_\nu-Z_\mu Z_\nu)+(X_\mu X_\mu+Y_\mu
Y_\mu)]+m^2\phi_\mu\phi_\nu
\]
where $k=(\lambda^2(1)+\lambda^2(2))$ for brevity. The first part
of $T_{\mu\nu}$, for either of the real fields is unaltered in
form when the tetrad is ``steered'' preserving the time-like
two-plane and the space-like two-plane. Therefore, \emph{without
loss of generality}, one may assume that $\phi_\mu$ takes one of
two forms \footnote{It is necessary to impose $\partial\cdot
\phi=0$ for solutions of the Proca equation.}. The algebraic
details are given in the appendix with the specification of the
tetrads.
\begin{enumerate}
    \item $\phi_\mu=\alpha T_\mu+\gamma X_\mu$
    \item $\phi_\mu=\beta Z_\mu+\delta Y_\mu$
\end{enumerate}
Taking an eigenvector $E_\mu$ in the form
\[
E_\mu=aT_\mu+bZ_\mu+cX_\mu+dY_\mu
\]
one gets, denoting the eigenvalue by $K$, two cases.

For case (i)
\begin{eqnarray}\nonumber
Ka&=&ka+m^2\alpha(\phi\cdot E)\\
\nonumber Kb&=&kb\\
\nonumber Kc&=&-kc+m^2\gamma(\phi\cdot E)\\
\nonumber Kd&=&-kd\\
\end{eqnarray}
and
\[
\phi\cdot E=a\alpha-c\gamma
\]
From the equations involving $a$ and $c$ only one gets
\begin{equation}\label{ev1}
\left(\frac{K-k}{K+k}\right)\frac{a}{c}=\frac{\alpha}{\gamma}
\end{equation}
One may take $a=\alpha(K+k)$ and $c=\gamma(K-k)$. The eigenvalue
equation is
\begin{equation}\label{ev2}(K-k-m^2\alpha^2)(K+k+m^2\gamma^2)+(m^2\alpha\gamma)^2=0
\end{equation}
The nature of the eigenvectors in this case is given by their norm
squared
\[
a^2-c^2=\alpha^2(K+k)-\gamma^2(K-k)=(K^2-k^2)\frac{1}{m^2}
\]
the last equality following from the equations (\ref{ev1},
\ref{ev2}). Since $(K-k-m^2\alpha^2)(K+k+m^2\gamma^2)$ is negative
one finds that
\begin{itemize}
    \item If $K$ is positive then $K<k+m^2\alpha^2$.
    \item If $K$ is negative then $K<-(k+m^2\gamma^2)$.
\end{itemize}
In this latter case $a^2-c^2$ is positive and $E_\mu$ will be
time-like. Since the two solutions are orthogonal the other solution
will be space-like. There are two further space-like solutions with
$\phi\cdot E=0$
\begin{itemize}
    \item $b$ only $c=d=0$, $K=+k$
    \item $d$ only $c=d=0$, $K=-k$
\end{itemize}
One has therefore four orthogonal solutions
\[
E_\mu=\alpha(K+k)T_\mu+\gamma(K-k)X_\mu
\]
(two solutions), and
\begin{eqnarray}
\nonumber E_\mu=Z_\mu\\
\nonumber E_\mu=Y_\mu\\
\end{eqnarray}

For case (ii)
\begin{eqnarray}\nonumber
Ka&=&ka\\
\nonumber Kb&=&kb+m^2\beta(\phi\cdot E)\\
\nonumber Kc&=&-kc\\
\nonumber Kd&=&-kd+m^2\delta(\phi\cdot E)
\end{eqnarray}
and
\[
\phi\cdot E=-(b\beta+d\delta)
\]
One immediately finds one time-like solution, with $K=+k$
\[
E_\mu=aT_\mu
\]
One space-like solution, with $K=-k$
\[E_\mu=cX_\mu
\]
and two space-like solutions (orthogonal to the first two
solutions).

\begin{equation}\label{Kk}
 \left(\frac{K-k}{K+k}\right)\frac{b}{d}=\frac{\beta}{\delta}
\end{equation}
One may take $b=\beta(K+k)$ and $d=\delta(K-k)$. The eigenvalue
equation is
\begin{equation}\label{ev equation2}
(K-k+m^2\beta^2)(K+k+m^2\delta^2)-(m^2\beta\delta)^2=0
\end{equation}
Again one has therefore four orthogonal solutions
\[\nonumber
E_\mu=\beta(K+k)Z_\mu+\delta(K-k)Y_\mu
\]
(two solutions), and
\begin{eqnarray}\label{Emu}
\nonumber E_\mu=T_\mu\\
\nonumber E_\mu=X_\mu\\
\end{eqnarray}
The overall eigenvalues of $\frac{T_{\mu\nu}}{2}$ from equation
(\ref{Tmunu}) will be
\[
\lambda=K-\frac{m^2}{2}\phi^\alpha\phi_\alpha
\]
For case (i)
\[
\lambda=\pm\sqrt{k^2+km^2(\alpha^2+\gamma^2)+(\frac{1}{2}m^2\phi^\alpha\phi_\alpha)^2}
\]
and
\[
\lambda=\pm k-\frac{m^2}{2}\phi^\alpha\phi_\alpha
\]
For case (ii)
\[
\lambda=\pm\sqrt{k^2+km^2(\delta^2-\beta^2)+(\frac{1}{2}m^2\phi^\alpha\phi_\alpha)^2}
\]
and
\[
\lambda=\pm k-\frac{m^2}{2}\phi^\alpha\phi_\alpha
\]
where
\begin{eqnarray}\nonumber
k&=&\frac{1}{4}\left[\left(G_{\alpha\beta}G^{\alpha\beta}\right)^2+\left(^*G_{\alpha\beta}
G^{\alpha\beta}\right)^2\right]^\frac{1}{2}\\
&=&\frac{1}{4}\left(^{(\theta)}G_{ab}\right)\left(^{(\theta)}G^{ab}\right)
\end{eqnarray}
where $\theta$ is the extremal angle of duality rotation (see
appendix). The eigenvectors of $T_{\mu\nu}$, including the mass
term, are as follows.

Case (i): The maxwellian part of $T_{\mu\nu}$ gives a time-like
two-plane $\left(\widehat{T}-\widehat{Z}\right)$ and the potential
$\phi_{\mu}$ is in the plane of
$\left(\widehat{T}-\widehat{X}\right)$, as shown in figure 1.
$\phi_\mu$ can be either time-like or space-like.

The time-like two-plane $\left(\widehat{T}-\widehat{X}\right)$
contains two eigenvectors which are Lorentz boosted $\widehat{T}$
and $\widehat{X}$ as can be easily seen from (\ref{Kk}),(\ref{ev
equation2}) and (\ref{Emu}).

Case (ii): Again the maxwellian part of $T_{\mu\nu}$ gives a
time-like two-plane $\left(\widehat{T}-\widehat{Z}\right)$ but
$\phi_{\mu}$ is in the plane of
$\left(\widehat{Z}-\widehat{Y}\right)$, as shown in figure 2. In
this case $E_\mu$ (time-like) is given by $\widehat{T}$ and
$\widehat{\phi}$ is perpendicular to $\widehat{T}$.

In both cases a definite time-like eigenvector is given; either
$\widehat{T}$ or a Lorentz-boosted version, the tetrads being
given in the appendix.
\section{Complex field and eigenvectors of $T_{\mu\nu}$}
In both cases the eigenvectors of $T_{\mu\nu}(i)$ enable one to
define a time-like two-plane. Just considering the parts of
$T_{\mu\nu}(i)$ defining the two time-like two-planes (omitting the
terms in the mass) one has
\begin{eqnarray}\label{complexTmunu}\nonumber
T_{\mu\nu}=&2k_1\left[T_\mu(1)T_\nu(1)-Z_\mu(1)Z_\mu(1)\right]\\
&+2k_2\left[T_\mu(2)T_\nu(2)-Z_\mu(2)Z_\mu(2)\right]
\end{eqnarray}
The $Z_{(1,2)}$ direction is a purely conventional label. The
common, unique, plane $\Omega$ with axes $\hat{t},\hat{z}$ is such
that \cite{Penrose}
\begin{enumerate}
\item $\widehat{Z}$ is orthogonal to
$\widehat{Z}(1)$,$\widehat{Z}(2$)
\item $T_\mu(i)=\cosh(\theta_i)t_\mu+\sinh(\theta_i)z_\mu$
 \end{enumerate}
The time-like eigenvector of the $T_{\mu\nu}$ given in
(\ref{complexTmunu}) will necessarily lie in the plane $\Omega$ and
take the form
\[T_\mu=\cosh(\theta)t_\mu+\sinh(\theta)z_\mu\]
$\Omega$ is unique but, of course, $\hat{t}$ and $\hat{z}$ may be
subjected to a Lorentz boost.

There also exists a space-like eigenvector in the $\Omega$ plane of
the form:
\[\sinh(\theta)t_\mu+\cosh(\theta)z_\mu\]
The eigenvalues given by
\[T_{\mu\nu}E^\nu=\Lambda E_\mu\]
are easily derived and are
\begin{enumerate}
\item Time-like eigenvector
\begin{eqnarray}\nonumber
\Lambda=&(k_1+k_2)\\
&+\left[\left(k_1\cosh(2\theta_1)+k_2\cosh(2\theta_2)\right)^2-\left(k_1\sinh(2\theta_1)+k_2\sinh(2\theta_2)\right)^2\right]^\frac{1}{2}\nonumber
\end{eqnarray}
\item Space-like eigenvector
\begin{eqnarray}\nonumber
\Lambda=&(k_1+k_2)\\
&-\left[\left(k_1\cosh(2\theta_1)+k_2\cosh(2\theta_2)\right)^2-\left(k_1\sinh(2\theta_1)+k_2\sinh(2\theta_2)\right)^2\right]^\frac{1}{2}\nonumber
\end{eqnarray}
\end{enumerate}
The three velocity derived from the time-like eigenvector is given
by $\tanh(\theta)$
\[\tanh(\theta)=\frac{\left[\left(k_1\cosh(2\theta_1)+k_2\cosh(2\theta_2)\right)\right]-\left[\Lambda-(k_1+k_2)\right]}{\left[\left(k_1\cosh(2\theta_1)+k_2\cosh(2\theta_2)\right)\right]+\left[\Lambda-(k_1+k_2)\right]}
\]
One can note that, in the massless case, the above results now give
a definite time-like eigenvector in the complex case.The terms
involving $Z_\mu(1)$ and $Z_\mu(2)$ can be similarly diagionalised.
$Z_\mu(1)$ and $Z_\mu(2)$ define a plane orthogonal to $\widehat{Z}$
so one can use definite unit orthogonal vectors
$\widehat{X},\widehat{Y}$ in this plane (in general $k_1\ne k_2)$.

An algebraic simplification can be brought about by applying a
lorentz boost to $(\hat{t},\hat{z})$ so that

\[
\theta_1'=-\theta_2'=\frac{\theta_1'-\theta_2'}{2}=\frac{\theta_1-\theta_2}{2}
\] and then
\[\tanh(2\theta')=\left(\frac{k_1-k_2}{k_1+k_2}\right)\tanh(\theta_1-\theta_2)\]
and
\[T_\mu(1)T^\mu(2)=\cosh(\theta_1-\theta_2)\]
(the dashed variables refer to the boosted axes). In the massive
vector field case the extra mass term, apart from the terms in
$g_{\mu\nu}$, will be
\[m^2\left(\phi_\mu(1)\phi_\nu(1)+\phi_\mu(2)\phi_\nu(2)\right)\]
In all cases mentioned in section (\ref{classification}) there will
now be cross terms between the $\left(\hat{t}-\hat{z}\right)$ plane
and the $\left(\hat{X}-\hat{Y}\right)$ plane defined for the
massless case.

One is now left with a complicated algebraic problem to find the
common time-like eigenvector of the total $T_{\mu\nu}$ which is left
for future investigations.

\section{An illustrative example}
We illustrate the flows of energy momentum using standing waves in
the $x_1-x_2$ plane and with spin $W^3$ equal to $+m$\footnote{ In
an earlier paper we used a similar example to calculate the particle
trajectories according to the de Broglie-Bohm interpretation of the
Dirac equation \cite{DHLMS}.}. We have previously calculated the
vector potential for this case in section (\ref{spind}) which gave
for the real and imaginary parts of $G_{\mu\nu}$,
$\left[G_{\mu\nu}\right]_R$ and $\left[G_{\mu\nu}\right]_I$
respectively

\[
[G_{\mu\nu}]_{R}= \left[ \begin{array}{cccc}
0 & mk_0c_2&-mk_0c_1&0 \\
-mk_0c_2 & 0&0&0\\
mk_0c_1 &0&0&0\\
0&0&0&0\\
\end{array}\right]
\]

\[
[G_{\mu\nu}]_{I}= \left[ \begin{array}{cccc}
0 & -(k_0^2-k_1^2)c_1&(k_0^2-k_2^2)c_2&0 \\
(k_0^2-k_1^2)c_1 & 0&m(k_1s_1+k_2s_2)&0\\
(k_0^2-k_2^2)c_2 &m(k_1s_1+k_2s_2)&0&0\\
0&0&0&0\\
\end{array}\right]
\]
where $c_2=cos(k_2x_2), c_1=cos(k_1x_1), s_2=sin(k_2x_2)$ and
$s_1=sin(k_1x_1)$. The null eigenvectors are easily calculated. For
the real field they are
\[
\left(%
\begin{array}{c}
  1 \\
  \pm\frac{\cos(k_2x_2)}{\sqrt{\cos^2(k_2x_2)+\cos^2(k_1x_1)}} \\
 \mp\frac{\cos(k_1x_1)}{\sqrt{\cos^2(k_2x_2)+\cos^2(k_1x_1)}}  \\
  0\\
\end{array}%
\right)
\]
This null dyad gives the time-like unit vector
\begin{equation}\label{TR}
\widehat{T}_R=
\left(%
\begin{array}{c}
  1 \\
0 \\
0 \\
  0\\
\end{array}%
\right)
\end{equation}
and the space-like unit vector
\begin{equation}\label{XR}
\widehat{X}_R=
\left(%
\begin{array}{c}\label{XR}
  0 \\
  \frac{\cos(k_2x_2)}{\sqrt{\cos^2(k_2x_2)+\cos^2(k_1x_1)}} \\
-\frac{\cos(k_1x_1)}{\sqrt{\cos^2(k_2x_2)+\cos^2(k_1x_1)}}  \\
  0\\
\end{array}%
\right)
\end{equation}
For the imaginary field the null vector is
\[
\left(%
\begin{array}{c}
  X^0 \\
 X^1 \\
X^2\\
  0\\
\end{array}%
\right)
\]
with
\[
\frac{X^1}{X^0}=\frac{-(\lambda A-\alpha B)}{(\lambda^2+\alpha^2)}
\]
and
\[
\frac{X^2}{X^0}=\frac{-(-\alpha A+\lambda
B)}{(\lambda^2+\alpha^2)}
\]
where
\begin{eqnarray}\label{AB}
\nonumber A=m^2\cos(k_1x_1)\\
\nonumber B=-m^2\cos(k_2x_2)\\
\nonumber \alpha=m(k_1\sin(k_1x_1)+k_2\sin(k_2x_2))\\
\lambda=\pm\sqrt{(A^2+B^2-\alpha^2)}\\
\end{eqnarray}\nonumber
which give normalized unit vectors for the time-like vector
\begin{equation}
\widehat{T}_I= \left(\begin{array}{c}
\frac{m^2}{\lambda}\sqrt{\cos^2(k_1x_1)+\cos^2(k_2x_2)}\\
\frac{\alpha}{\lambda}\frac{\cos(k_2x_2)}{\sqrt{\cos^2(k_2x_2)+\cos^2(k_1x_1)}}\\
-\frac{\alpha}{\lambda}\frac{\cos(k_1x_1)}{\sqrt{\cos^2(k_2x_2)+\cos^2(k_1x_1)}}\\
0\\
\end{array}
\right)
\end{equation}
and the space-like vector
\begin{equation}\label{XI}
\widehat{X}_I= \left(
\begin{array}{c}
0\\
\frac{\cos(k_1x_1)}{\sqrt{\cos^2(k_2x_2)+\cos^2(k_1x_1)}}\\
\frac{\cos(k_2x_2)}{\sqrt{\cos^2(k_2x_2)+\cos^2(k_1x_1)}}\\
0\\
\end{array}
\right)
\end{equation}
A complication arises at points where $\lambda$ goes from real to
imaginary. In the latter case the two null eigenvectors of
$G_{\mu\nu}$ give rise to two space-like vectors. The vector
orthogonal to the vector that goes through from time-like to
space-like, and thereafter gives the unit time-like vector, is
\begin{equation}\label{tvec}\left(
\begin{array}{c}
\frac{\alpha}{|\lambda|}\\
\frac{m^2}{|\lambda|}\cos(k_2x_2)\\
\frac{-m^2}{|\lambda|}\cos(k_1x_1)\\
0\\
\end{array}\right)
\end{equation}
This is proportional to $\phi_\mu$ (imaginary field) which switches
over in the same way. We now have two time-like two-planes as shown
in figure 3. One notes that $\widehat{T}_{I}$ is obtained by a
Lorentz boost in the plane of $(\widehat{T}_R,\widehat{X}_R)$ so
that
\[
\widehat{T}_{I}=\cosh(\theta)\widehat{T}_R+\sinh(\theta)\widehat{X}_R
\]
therefore
\begin{eqnarray}\nonumber
\cosh\theta&=&\frac{m^2}{\lambda}\sqrt{\cos^2(k_1x_1)+\cos^2(k_2x_2)}\\
\nonumber \sinh\theta&=&\frac{\alpha}{\lambda}\\
\nonumber
\tanh\theta&=&\frac{\alpha}{m^2}\sqrt{\cos^2(k_1x_1)+\cos^2(k_2x_2)}\\
\end{eqnarray}
In this case the intersection of the two time-like two-planes gives
the common vector $\widehat{T}_{I}$ so that one can express both
parts of $T_{\mu\nu}$ in terms of $\widehat{T}_{I}$ and
$\widehat{X}_R$ boosted.

\subsection{Derivation of the Time-like Eigenvector} Apart from
terms in $g_{\mu\nu}$, $T_{\mu\nu}$ can be written as (see section
\ref{secrealT})
\begin{eqnarray}\nonumber
k_R\left[T_{\mu R}T_{\nu R}-X_{\mu R}X_{\nu R}\right]+m^2\phi_{\mu
R}\phi_{\nu
R}\\
\nonumber +k_I\left[T_{\mu I}T_{\nu I}-X_{\mu I}X_{\nu
I}\right]+m^2\phi_{\mu I}\phi_{\nu I}\\ \label{exp}
\end{eqnarray}
Without changing the form of the expression one may apply a Lorentz
boost to $(T_{\mu R},X_{\mu R})$. Noting that $\widehat{X}_R$ is
orthogonal to $\widehat{X}_I$ (\ref{XR},\ref{XI}) we change our
notation as follows
\begin{eqnarray}
\nonumber \widehat{T}_{\mu I}\longrightarrow\widehat{T}_\mu\\
\nonumber \widehat{X}_{\mu I}\longrightarrow\widehat{Y}_\mu\\
\nonumber \widehat{X}_R(boosted)\longrightarrow\widehat{X}
\end{eqnarray}
the relationship of the vectors is shown in figure 4. We can now
write the expression (\ref{exp}) as
\begin{eqnarray}\nonumber
k_R\left[\widehat{T}_{\mu}\widehat{T}_{\nu}-\widehat{X}_{\mu}\widehat{X}_{\nu}\right]+m^2\phi_{\mu
R}\phi_{\nu
R}\\
\nonumber
+k_I\left[\widehat{T}_{\mu}\widehat{T}_{\nu}-\widehat{Y}_{\mu}\widehat{Y}_{\nu}\right]+m^2\phi_{\mu
I}\phi_{\nu
I}\\
\end{eqnarray}
$\phi_{\mu R}$ is orthogonal to both $\widehat{T}_R$ and
$\widehat{X}_R$ and proportional to
$\widehat{X}_I\equiv\widehat{Y}_R$ (which we have now labelled
$\widehat{Y}$). $\phi_{\mu I}$ is orthogonal to both
$\widehat{T}_{\mu I} (\equiv \widehat{T}_\mu)$ and $\widehat{X}_{\mu
I} (\equiv \widehat{Y}_\mu)$ and is now proportional to the boosted
$\widehat{X}_R$, which is now $\widehat{X}$. We now have
\begin{eqnarray}\nonumber
k_R\left[\widehat{T}_{\mu}\widehat{T}_{\nu}-\widehat{X}_{\mu}\widehat{X}_{\nu}\right]+m^2\widehat{Y}_{\mu}\widehat{Y}_{\nu}\times(const)\\
\nonumber +k_I\left[\widehat{T}_{\mu}\widehat{T}_{\nu}-\widehat{Y}_{\mu}\widehat{Y}_{\nu}\right]+m^2\widehat{X}_{\mu}\widehat{X}_{\nu}\times(const)\\
\end{eqnarray}
One immediately sees that $\widehat{T}$ is the time-like eigen
vector. The four velocities are then
\begin{eqnarray}
\nonumber
\frac{dt}{d\tau}=\frac{m^2}{\lambda}\sqrt{\cos^2(k_1x_1)+\cos^2(k_2x_2)}\\
\nonumber
\frac{dx_1}{d\tau}=\frac{\alpha}{\lambda}\frac{\cos(k_2x_2)}{\sqrt{\cos^2(k_1x_1)+\cos^2(k_2x_2)}}\\
\nonumber
\frac{dx_2}{d\tau}=-\frac{\alpha}{\lambda}\frac{\cos(k_1x_1)}{\sqrt{\cos^2(k_1x_1)+\cos^2(k_2x_2)}}\\
\end{eqnarray}
with definitions as given previously in equation (\ref{AB}).

The eigenvalue belonging to the time-like eigenvector is easily
calculated from the $G_{\mu\nu}$ and $\phi_\mu$ previously given
(see section \ref{spind}). In this particular example the terms in
$g_{\mu\nu}$ are equal to zero and therefore the overall eigenvalue
will be
\begin{eqnarray}\label{oev}
k_R+k_I&=&\left[m^2k_0^2\left(cos^2(k_1x_1)+cos^2(k_2x_2)\right)\right]\\\nonumber
&+&\left[m^4\left(cos^2(k_1x_1)+cos^2(k_2x_2)\right)-m^2\left(k_1
sin(k_1x_1)+k_2 sin(k_2x_2)\right)^2\right]
\end{eqnarray}
and $|k_1|=|k_2|$.

When $\lambda$ becomes imaginary the four velocities can be read
off from the new unit time-like vector (\ref{tvec}) as
\begin{eqnarray}\nonumber
\frac{dt}{d\tau}=\frac{\alpha}{|\lambda|}\\
\frac{dx_1}{d\tau}=m^2\frac{\cos(k_2x_2)}{|\lambda|}\\\nonumber
\frac{dx_2}{d\tau}=-m^2\frac{\cos(k_1x_1)}{|\lambda|}\\\nonumber
\end{eqnarray}

The overall eigenvalue remains as given before in equation
(\ref{oev}). The standing wave pattern which arises in the
eigenvalue is shown in figure 5. For the purposes of the
illustration we take $m=\hbar=c=1$ and $k_1=k_2=0.2$. Figure 6 shows
the flow-lines in $x_1,x_2,t$ around one of the minima in the
eigenvalue.

\section{Appendix}
A general result for choosing tetrads in the case of the
maxwellian field has been given in \cite{Ruse1936},
\cite{Misner1957} and \cite{Garat2004}. In the case of current
interest one needs to include the specification of the vector
potential $\phi_\mu$ by the tetrads. We therefore need to modify
the choices made in \cite{Ruse1936}, \cite{Misner1957} and
\cite{Garat2004}.

The vector field tensor, $G_{\mu\nu}$, is subjected to a duality
rotation \cite{Penrose} giving an extremal field
\[
^{(\theta)}G_{\mu\nu}=G_{\mu\nu}cos(\theta)+^*G_{\mu\nu}sin(\theta)
\]
One finds that
\[
\left(^{(\theta)}G_{\mu\nu}\right)\left(^{*(\theta)}G^{\nu\lambda}\right)=g_\mu^{
\lambda}\left[cos(2\theta)(\underline{E}\cdot\underline{B})-sin(2\theta)\left(\frac{E^2-B^2}{2}\right)\right]
\]
where \underline{E} and \underline{B} are the two 3-vectors
specifying $G_{\mu\nu}$. Choosing
\[
tan(2\theta)=\frac{2E\cdot B}{E^2-B^2}
\]
gives \begin{equation}\label{extremal}
\left(^{(\theta)}G_{\mu\nu}\right)\left(^*(\theta)G^{\nu\lambda}\right)=0
\end{equation}
One notes from \cite{Ruse1936} and \cite{Misner1957}, that the
maxwellian part of the stress-energy-momentum tensor is unaltered
by such a duality rotation. In terms of the extremal field and two
arbitrary vectors one obtains a tetrad defining a time-like
two-plane and an orthogonal space-like two-plane:
\begin{eqnarray}
\nonumber U_\mu=k\left(^{(\theta)}G_{\mu\nu}\right)\theta^\nu\\
\nonumber V_\mu=\frac{1}{k}\left(^{(\theta)}G_{\mu\nu}\right)U^\nu\\
\nonumber W_\mu=k\left(^*\left(^{(\theta)}G_{\mu\nu}\right)\right)\triangle^\nu\\
\nonumber Z_\mu=\frac{1}{k}\left(^*\left(^{(\theta)}G_{\mu\nu}\right)\right)W^\nu\\
\end{eqnarray}
$\theta^\nu$ and $\triangle^\nu$ are two arbitrary vectors, in
general, but will be chosen to be $\phi^\nu$ for our purposes. k
is the magnitude of the eigenvalues of the maxwellian part of the
stress-energy-momentum tensor (its inclusion is not necessary for
the arguments following but is helpful in normalisation.) The
tetrads are not normalised as given. The extremal property
(\ref{extremal}) ensures that the two-planes defined by
$\left(U_\mu,V_\mu\right)$ and $\left(W_\mu,Z_\mu\right)$ are
orthogonal for any choice of $\theta^\nu$ and $\triangle^\nu$.
Using the anti-symmetry of $^{(\theta)}G_{\mu\nu}$ it is easy to
show that: \begin{enumerate}
    \item $\theta^\nu$ is orthogonal to $U^\nu$
    \item $V^\nu$ is orthogonal to $U^\nu$
\end{enumerate} A similar result will also apply to the other
two-plane involving the duals of $^{(\theta)}G_{\mu\nu}$. The
plane defined by $\left(U_\nu,V_\nu\right)$ will be time-like (see
the details in section \ref{secrealT}). Choosing
$\theta^\nu=\phi^\nu$ one has
\begin{enumerate}
    \item $\phi_\nu$ time-like gives $V^\nu$ space-like and hence
    $U^\nu$ time-like
    \item $\phi^\nu$ space-like gives $V^\nu$ time0like and hence
    $U^\nu$ space-like
\end{enumerate}
Using the orthogonality properties of $\theta^\nu$ and
$\triangle^\nu$ one finds that $\phi^\nu$ has components in both
two-planes
\[
\phi^\nu=\frac{\left(\phi\cdot U\right)}{\left(U\cdot
U\right)}U^\nu+\frac{\left(\phi\cdot Z\right)}{\left(Z\cdot
Z\right)}Z^\nu
\]
or
\[
\phi^\nu=\frac{\left(\phi\cdot V\right)}{\left(V\cdot
V\right)}V^\nu+\frac{\left(\phi\cdot W\right)}{\left(W\cdot
W\right)}W^\nu
\]
(This confirms the intuitively obvious result used in section
\ref{secrealT}).
\newpage
\section{Figure Captions}
\begin{description}
  \item[Figure 1] The maxwellian part of $T_{\mu\nu}$ gives a time-like
two-plane $\left(\widehat{T}-\widehat{Z}\right)$ and the potential
$\phi_{\mu}$ is in the plane of
$\left(\widehat{T}-\widehat{X}\right)$
  \item[Figure 2] The maxwellian part of $T_{\mu\nu}$ gives a
time-like two-plane $\left(\widehat{T}-\widehat{Z}\right)$ but
$\phi_{\mu}$ is in the plane of
$\left(\widehat{Z}-\widehat{Y}\right)$
  \item[Figure 3] The intersection of the two time-like two-planes
  gives the common vector $\widehat{T}_{Imaginary}$
  \item[Figure 4] The relationship
between the vectors, $\widehat{T}_R,\widehat{T}, \widehat{X},
\widehat{X}_R, \widehat{Y}$.
  \item[Figure 5] The standing wave pattern, in the $x_1-x_2$ plane, which arises in the
eigenvalue for the illustrative example ($m=\hbar=c=1$ and
$k_1=k_2=0.2$).
 \item[Figure 6] A set of flow lines of energy-momentum in the $x_1-x_2-t$
space-time for the illustrative example ($m=\hbar=c=1$ and
$k_1=k_2=0.2$).
\end{description}

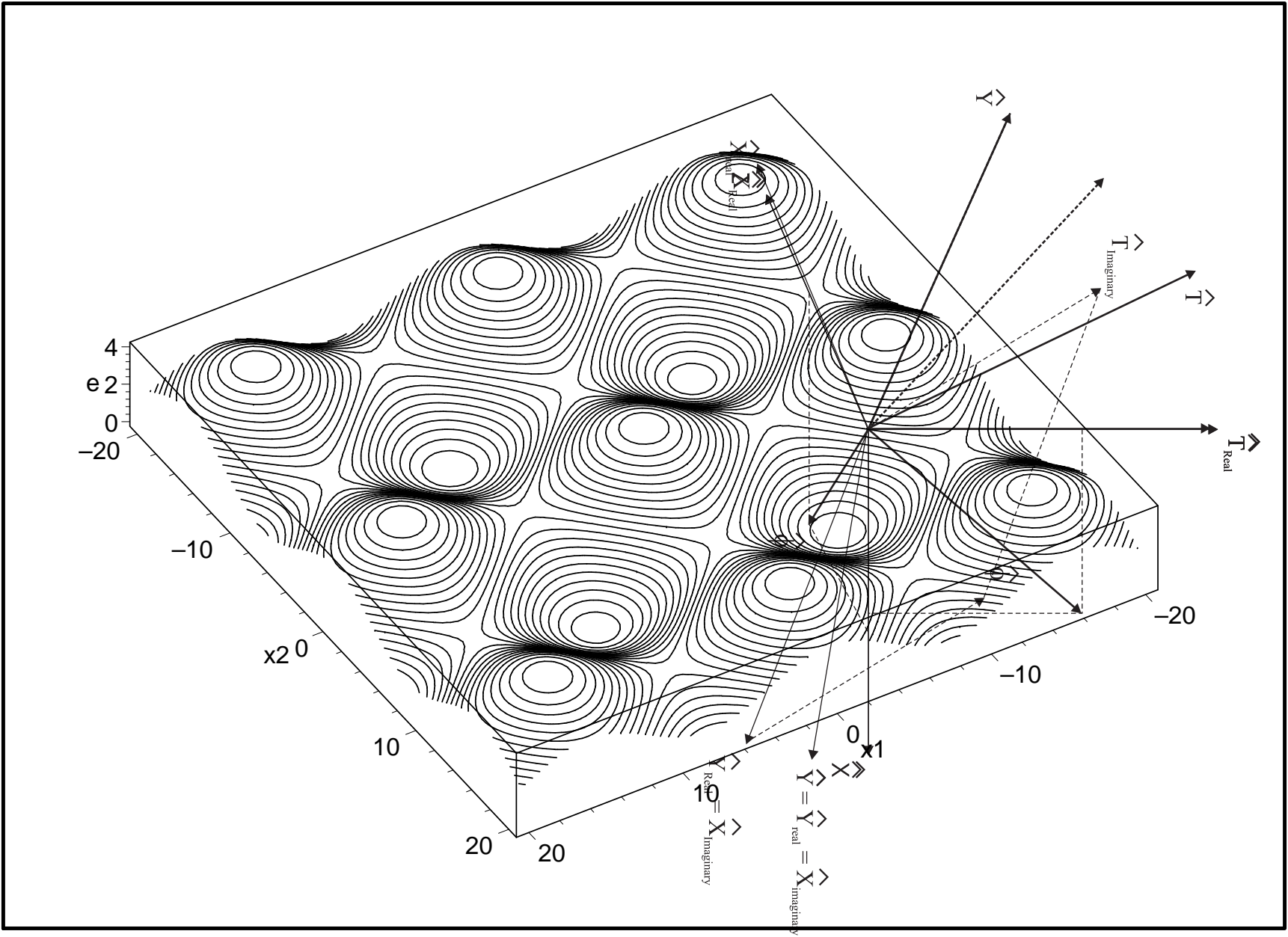
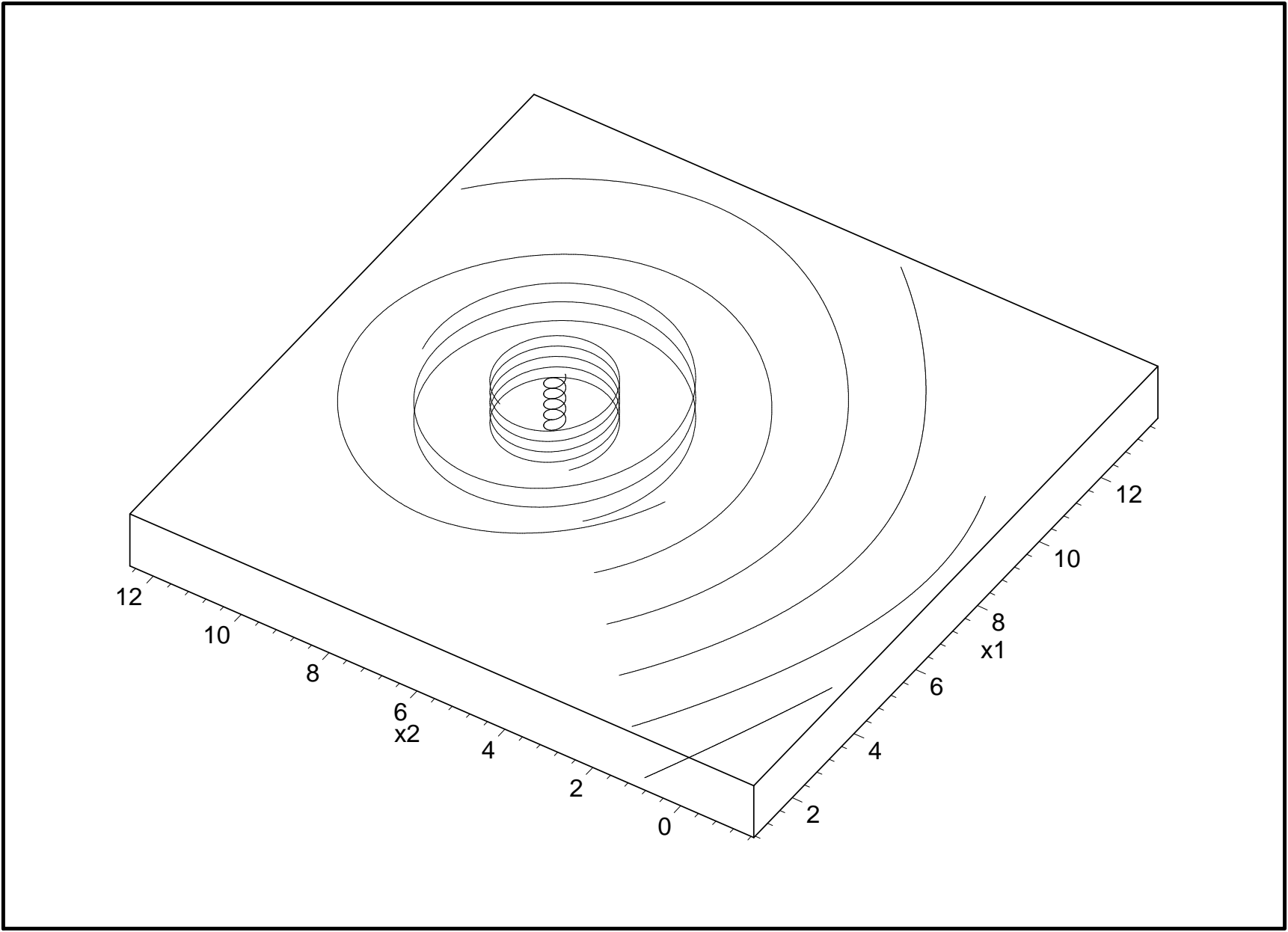
\end{document}